## Title

**Interaction-driven flat band and charge order in $Fe_5GeTe_2$**

## Authors

Qiang Gao[1,#], Gabriele Berruto[1,#], Khanh Duy Nguyen[1], Chaowei Hu[2], Paul Malinowski[2], Haoran Lin[1], Beomjoon Goh[3], Bo Gyu Jang[4], Xiaodong Xu[2], Peter Littlewood[5,6], Jiun-Haw Chu[2], and Shuolong Yang[1,*]

## Affiliations

[1]Pritzker School of Molecular Engineering, The University of Chicago, Chicago, Illinois 60637, USA

[2]Department of Physics, University of Washington, Seattle, Washington 98195, USA

[3]Department of Physics and Astronomy, Seoul National University, Seoul 08826, Republic of Korea

[4]Department of Advanced Materials Engineering for Information and Electronics, Kyung Hee University, Yongin 17104, Republic of Korea

[5]James Franck Institute and Department of Physics, The University of Chicago, Chicago, IL 60637, USA

[6]School of Physics and Astronomy, University of St Andrews, N Haugh, St Andrews KY16 9SS, United Kingdom

[#]These authors contributed equally to this work.

*Corresponding author. Email: yangsl@uchicago.edu

## Abstract

Flat electronic bands enable fascinating emergent phenomena such as superconductivity and charge orders. A prevailing approach to realizing flat bands is to engineer lattice geometric constraints in twisted or kagome-like materials. An alternative approach is to utilize purely electronic-interaction-driven flat bands, yet a fundamental challenge is that extreme flatness requires ultrastrong interaction strength, which often leads to incoherent states. Here we demonstrate the concurrent formation of an interaction-driven flat band at the Fermi level and a $\sqrt{3} \times \sqrt{3}R30°$ charge order in a van der Waals magnet $Fe_5GeTe_2$ using high-resolution angle-resolved photoemission spectroscopy. This charge order is manifested by band folding within 30 meV below the Fermi level, with its nesting driven by flat bands. The presence of this flat band throughout the Brillouin zone and the logarithmic temperature dependence of its spectral weight suggest a Kondo-like, coherent Fermi liquid emerging from strong correlations. Our work establishes a paradigm where an interaction-driven flat band promotes large-scale electronic ordering.

## MAIN TEXT

### Introduction

Flat electronic bands in solid-state systems can fundamentally reshape the collective ground states and lead to fascinating orders such as charge orders (*1*), superconductivity (*2*, *3*), correlated insulators (*4*), as well as integer (*5*) and fractional topological orders (*6*-*8*). Flat bands can be obtained by engineering lattice geometric constraints, most notably via moiré potentials in twisted 2D materials and via destructive quantum interference in special lattices. Yet, in twisted 2D materials, the flat-band condition is often unstable against sample handling, and the uniformity of the moiré pattern is restricted to < 1 μm. In lattices with special geometries, such as Lieb (*9*, *10*), kagome (*11*-*15*), and clover (*16*) lattices, the flat bands are often hundreds of meV away from the Fermi level ($E_F$), making them irrelevant for low-energy ordering phenomena. Moreover, such flat bands are fully determined by the crystal structures and exhibit minimum tunability. Therefore, there has been a growing interest in finding alternative flat-band systems that are robust, tunable, and extendable to large spatial scales, ideal for investigating the interplay of electron correlations, topology, and charge orders at $E_F$ (*17*, *18*).

Electronic-interaction-driven flat bands provide a different possibility. Strong electron-electron, electron-phonon, or electron-spin interactions lead to band structure renormalization close to $E_F$ (*19*). In fact, in some twisted 2D materials (*20*) or geometrically constrained lattices (*21*), electronic interactions have been shown to contribute significantly to band flattening. It is thus inspiring to explore purely interaction-driven flat bands for emergent quantum phenomena. Such band flattening can be robust against sample handling, tunable by carrier doping or temperature variation, and extendable to macroscopic length scales. However, a fundamental challenge is that extreme band flatness often requires a diverging interaction strength, which in turn can lead to incoherent states. Such incoherent interaction-driven flat bands (*22*) cannot be utilized to enable emergent orders near $E_F$.

Here we focus on $Fe_5GeTe_2$ which is a van der Waals magnetic material holding an exciting potential for flat-band engineering. $Fe_5GeTe_2$ has multiple structural phases distinguished by the ordering of the Fe(1) atoms, which can occupy one of the up and down sites relative to the Ge atom in each unit cell (*23*-*28*). In the so-called site-ordered phase, the Fe(1) atoms follow an "up-down-down" or "down-up-up" (UDD or DUU) arrangement to form a clover lattice (Fig. 1B), where flat bands away from $E_F$ emerge due to destructive quantum interference. Meanwhile, theory predicted that an effective Kondo-like interaction between the flat bands and the dispersive conduction bands in a clover lattice leads to new flat bands exactly at $E_F$ (*16*). A recent scanning tunneling microscopy (STM) study (*29*) revealed that flat bands appear concomitantly with a $\sqrt{3} \times \sqrt{3}R30°$ charge order in the "UUU" phase of $Fe_5GeTe_2$. The crucial difference between this charge order and the structural $\sqrt{3} \times \sqrt{3}$ order in the UDD or DUU phase is that the former does not exhibit the lattice reconstruction and is purely electronically driven. However, so far flat bands at $E_F$ associated with either the electronic or structural $\sqrt{3} \times \sqrt{3}$ order in $Fe_5GeTe_2$ have not been clearly revealed using angle-resolved photoemission spectroscopy (ARPES) (*28*, *30*-*33*), making it difficult to investigate how flat bands impact electronic orders in this van der Waals magnet. We note that recent studies on $Fe_5GeTe_2$ in the slow-cooled phase (*34, 35*) reported flat bands close to $E_F$, yet no Fermi surface reconstruction due to a $\sqrt{3} \times \sqrt{3}$ order has been reported, making it not suitable for studying the interplay between flat bands and charge orders.

In this article we report the discovery of a $\sqrt{3} \times \sqrt{3}R30°$ charge order driven by flat bands at $E_F$ in $Fe_5GeTe_2$. Multiple phases are identified by high-resolution ARPES. In one phase unrevealed by previous ARPES studies, electronic structures are folded to the $\sqrt{3} \times \sqrt{3}$ reconstructed Brillouin zone, yet the band folding is limited to a small binding energy range less than 30 meV, indicating the purely electronic origin of this order. The nesting vector connects segments of the Fermi surfaces where the flat band is pronounced in the ARPES spectrum. The characteristics of the flat band, that it is omnipresent throughout the Brillouin zone and exhibits a logarithmically growing spectral weight towards lower temperatures, point to a Kondo-like, coherent Fermi liquid emerging from strong correlation physics. Taken together with the calculation of Lindhard response functions for both the dispersive-band and flat-band regimes, we conclude that the flat band dramatically enhances the tendency for Fermi surface nesting. This makes $Fe_5GeTe_2$ a unique non-twisting system to demonstrate the generalizable scheme where interaction-driven flat bands at $E_F$ lead to electronic ordering.

**Results**

The electronic structures of two distinct phases of $Fe_5GeTe_2$ are revealed by ARPES using 21.2 eV photons (Fig. 1). Importantly, we use 6 eV laser-based micrometer-scale ARPES to visualize the microscopic mixture of different phases (fig. S1). This enables us to select the material allowing helium-lamp-based ARPES at 21.2 eV with a 0.5 mm beam spot to investigate regions dominated by a single phase. By rapidly quenching the sample from 550 K to 273 K we preferentially obtain Phase I (Fig. 1E-G). Here, a hexagonal hole pocket and a quasi-circular electron pocket are observed at the $\bar{\Gamma}$ and $\bar{\mathrm{K}}$ points of the unreconstructed Brillouin zone. These observations are generally consistent with those in the previously revealed site-ordered phase (*28*). Moreover, a geometrically induced flat band at -0.5 eV is observed in this phase (fig. S2), reinforcing the assignment to the site-ordered phase (*28*). Consistent with the prior study on the site-ordered phase (*28*), we also do not observe band folding in this phase, possibly due to the short coherence length of each "UDD" or "DUU" domain (*29*). By slowly cooling the sample from 550 K to 300 K with a 2-4 K/min cooling rate we have a higher probability of observing Phase II, where diffusive features are observed at the $\bar{\Gamma}$ and $\bar{\mathrm{K}}$ points in the Fermi surface map (Fig. 1H). 6 triangular features are resolved at the 6 corners of the reconstructed Brillouin zone. The Fermi surface topology is consistent with a $\sqrt{3} \times \sqrt{3}$ reconstructed Brillouin zone. The rather diffusive features in the Fermi surface map reflect flat band dispersions in the energy-momentum cuts (Fig. 2D). We also reveal a more subtle band folding at the binding energy of 30 meV (Fig. 1I), where the original hexagonal contour at $\bar{\Gamma}$ leads to folded replicas at the 6 $\bar{\mathrm{K}}$ points. The low-energy electronic ordering is also manifested in the replication of the band dispersions at $\bar{\Gamma}$ and $\bar{\mathrm{K}}$ (Fig. 2C). Finally, we do not observe band folding at the binding energy of 80 meV (Fig. 1J). The confined energy range for band folding is inconsistent with a lattice-based $\sqrt{3} \times \sqrt{3}$ order, indicating that the charge order in Phase II is likely electronically driven.

A flat band at $E_F$ is observed in the charge-ordered Phase II. This is seen by plotting the energy-momentum cuts along the high-symmetry trajectories in the momentum space (Fig. 2B). Figure 2D shows that the Fermi surface feature at $\bar{\Gamma}$ and the triangular features at the folded zone corners $\bar{\mathrm{K}}'$ are related to flat band dispersions ($\alpha$). To better visualize the $\alpha$ band, the energy-momentum spectra (Fig. 2D, F, G) are divided by the Fermi-Dirac function convolved with the energy resolution. Raw spectra are plotted in fig. S4. The energy distribution curves (EDCs) taken around $\bar{\Gamma}$ and $\bar{\mathrm{K}}'$ exhibit a pronounced peak which is relatively momentum-independent and pinned at $E_F$ (Fig. 2E), demonstrating a non-dispersive band at $E_F$. In addition, a weakly-dispersive, electron-like band ($\beta$) at 30 meV below $E_F$ is observed by both 21.2 eV (Fig. 2D) and 6 eV photons with a *p*-polarization (Fig. 2F). *S*-polarized 6 eV light preferentially enhances the $\alpha$ band over a broader momentum range (Fig. 2G, H). Figure 2G also suggests that the $\alpha$ feature connects to a dispersive hole-like band at momenta beyond [-0.2, 0.2] $Å^{-1}$. The *p*- and *s*-polarized light yield stronger cross-sections on the right and left part of the electronic structure, respectively. A more detailed study suggests that the linear dichroism follows the $C_3$ symmetry (fig. S5), as expected for the rhombohedral stacking of $Fe_5GeTe_2$ slabs (*23*). In a traditional electronically driven charge-density wave, the tendency for nesting is the strongest when there is a nesting vector connecting parallel segments of the Fermi surface (*36*). However, such a nesting vector defined by the hexagonal pocket (solid orange in Fig. 2A) is along $\bar{\Gamma} - \bar{\mathrm{M}}$, inconsistent with the folding between $\bar{\Gamma}$ and $\bar{\mathrm{K}}$. Instead, the nesting vector directly connecting $\bar{\Gamma}$ and $\bar{\mathrm{K}}$ links the momentum-space segments where the flat band feature $\alpha$ is pronounced. We will show in the Discussion section that such a flat band configuration dramatically promotes electronic ordering. Moreover, Fig. 2D and Fig. 2F illustrate that flat dispersions are seen both at 0 and 30 meV below $E_F$, which explains the fact that band folding at these energies is detectable, as shown in Fig. 1H and 1I.

We discuss the possible structural origin of Phase II. Importantly, structural phase assignments of $Fe_5GeTe_2$ in the ARPES literature (*28*, *30-35*) have been particularly controversial due to the multiple possible structural configurations overlooked by early ARPES studies. Here we mainly compare our results with a recent study (*28*) considering its explicit investigation of the lattice symmetry using optical second harmonic generation. A more comprehensive comparison with all ARPES studies on $Fe_5GeTe_2$ is included in the supplementary text. Rapid thermal quenching and slow cooling of $Fe_5GeTe_2$ have been reported to lead to the site-ordered and site-disordered phases, respectively (*28*). In the latter phase, the up and down Fe(1) sites are randomly occupied in each unit cell. While it is tempting to attribute our Phase II to the site-disordered phase according to the thermal trajectory, Phase II exhibits a markedly different electronic structure. The site-disordered phase features multiple dispersive hole pockets near $\bar{\Gamma}$ and dispersive electron pockets near $\bar{\mathrm{K}}$. Neither the $\sqrt{3} \times \sqrt{3}$ Fermi surface reconstruction nor the flat band at $E_F$ was observed in the site-disordered phase (*28*). Moreover, we utilized 6 eV laser-based micro-ARPES and discovered two kinds of electronic structures different from that of the site-ordered phase (fig. S3). One corresponds to Phase II where diffusive Fermi surface features appear at $\bar{\Gamma}$ and $\bar{\mathrm{K}}$ associated with flat band dispersions (Fig. 2A). The other (Phase III) shows dispersive hole bands near $\bar{\Gamma}$ but no strong feature at -0.5 eV. The feature at -0.5 eV is a signature of the site-ordered phase (fig. S1B). This comparison suggests that Phase III is more consistent with the characteristics of the site-disordered phase, but Phase II is yet a different phase. Considering the observation of the $\sqrt{3} \times \sqrt{3}$ Fermi surface reconstruction and that of the flat bands, we conclude that the most likely scenario is that Phase II corresponds to the charge-ordered state identified with the “UUU” configuration in a previous STM study (*29*).

We proceed to reveal the nature of the flat band at $E_F$ by conducting temperature-dependent measurements (Fig. 3). For this study we primarily use 6 eV laser-based ARPES with ~10 μm spatial resolution. Raw spectra are divided by the Fermi-Dirac function convolved with the energy resolution. This analysis shows the band dispersion above $E_F$ within a limited energy range. Below 100 K, $\alpha$ and $\beta$ bands can be distinguished (Fig. 3A-D). Above 100 K, the two bands seemingly merge into a hole pocket with the band top ~ 10 meV above $E_F$ (Fig. 3F). Our observations do not exclude the possibility that the broadening of the $\alpha$ and $\beta$ bands at high temperatures leads to the appearance of “merging,” yet they remain as separate bands throughout the measurement temperature range. Additional Helium-lamp-based ARPES measurements indicate that the band structure near $E_F$ should be best interpreted as intercepting electron and hole bands (fig. S6). Near the Fermi crossings of the electron and hole bands, dramatic band flattening appears at low temperatures (Fig. 3G). In Fig. 3K, we plot the temperature-dependent spectral weight of the $\alpha$ feature, integrated in the range of $[-50, 0]$ meV, after subtracting the incoherent background modeled by a Fermi-Dirac function. In addition, we plot the spectral height of the $\alpha$ feature after subtracting the Fermi-Dirac background. Both quantities scale linearly with $-\log(T)$, where $T$ is the temperature.

These observations allow us to examine various theoretical scenarios for the formation of the flat band. First, strong electron-phonon coupling can lead to renormalized band dispersions near $E_F$. The extreme case is the formation of polarons, which give rise to peak-dip-hump features with the dip energy close to the phonon energy (*37-39*). Under this scenario, we may associate the spectral dip between $\alpha$ and $\beta$ (Fig. 3A) with phonons near 15 meV (*40*, *41*). However, such a modification of the single-particle spectral function near $E_F$ should conserve the integrated spectral weight (fig. S7), which is not the case in our results when integrating the background-subtracted spectral weight in the range of either $[-50, 0]$ meV (Fig. 3K) or $[-230, 80]$ meV (fig. S8E). In addition, even with an artificially high electron-phonon coupling constant of $\lambda = 1.0$, the polaronic band structure (fig. S7) still does not yield the extremely flat band $\alpha$ (Fig. 2D), rendering this scenario unlikely. Another possibility is the formation of excitonic band structure, given that the $\sqrt{3} \times \sqrt{3}$ charge order folds an electron pocket from $\overline{\mathrm{K}}$ to $\overline{\Gamma}$ to hybridize with the hole pocket at $\overline{\Gamma}$. In the weak-coupling limit, the lower branch of the hybridized band appears in an "M" shape, but in the strong-coupling limit it can have a flat top (fig. S9) (*42-44*). However, this scenario does not explain the formation of the flat band also near the $\overline{\mathrm{K}'}$ point (Fig. 2D). In addition, $Fe_5GeTe_2$ is metallic (*24*) where excitonic instabilities should in general be screened. We also considered a scenario where spin-dependent scattering leads to flat dispersions near $E_F$, yet the spectral weight and linewidth of the flat band are calculated to be inconsistent with the log $(T)$ temperature scaling (fig. S10). Motivated by the previous STM study (*29*), we consider a Kondo-like mechanism. The $\log(T)$ scaling of the coherent spectral weight and of the spectral height is consistent with the Kondo-like mechanism (*45*, *46*). Moreover, the temperature-dependent FWHM of the $\alpha$ band agrees with the theoretically predicted linewidth of an effective Kondo resonance peak (fig. S8D), $w(T) = 2\sqrt{(\pi k_B T)^2 + 2(k_B T_K)^2}$, where $T$ and $T_K$ are the sample temperature and the single-ion Kondo temperature; $k_B$ is the Boltzmann constant (*47-49*). However, this comparison does not allow us to accurately determine $T_K$, as the theory-experiment agreement is not obviously better for any specific $T_K$ value (fig. S8D). Our experimental linewidth determined from ARPES is overall smaller than what reported by fitting the Fano line shape in STM in the comparable temperature range (*29*). We speculate that this is because STM probes the momentum-integrated density of states, while micro-ARPES allows us to focus on a small momentum region which helps exclude contributions from other momentum regions without pronounced flat bands.

## Discussion

The interaction-driven flat band in $Fe_5GeTe_2$ is highly intriguing. Even though Kondo physics is traditionally associated with *f*-electrons, Kondo-like systems with *d*-electrons have been identified, such as $Fe_3GeTe_2$ (*50*) and $YFe_2Ge_2$ (*51*). A fundamental conceptual difficulty is how the extended *d*-electron wavefunctions enable Kondo-like interactions. It was argued by Xu et al. (*51*) that the kinetic frustration, band hybridization, and electronic correlations for *d*-electrons can lead to quasi-localized states, which play the role of the localized states in *f*-electron systems. One should note that such a scenario is by no means equivalent to traditional Kondo lattices, as the coherent quasiparticle peak and hybridization gap do not vanish entirely above the coherence temperature ($T^*$) (*51*). The *d*-electrons are also not fully localized at high temperatures, but rather exhibit a coherent-to-incoherent crossover as shown in our Fig. 3. Such a coherent-to-incoherent crossover can emerge from a generic Hubbard model with a high coordinate number, where no explicit localized state is required but a heavily renormalized Fermi-liquid-like state emerges due to strong correlation effect (*52-54*). However, the observable quantities, such as extremely flattened band dispersions at $E_F$, can be effectively described by a Kondo-like interaction, which was shown theoretically (*53*). Lacking a standard terminology to describe this strong correlation effect, the word "Kondo" here is used to only represent the interaction-driven flat band exhibiting Kondo-like temperature scaling.

The discovery of interaction-driven, Kondo-like flat bands helps us understand unusual characteristics of $Fe_5GeTe_2$ by electrical transport and magnetization measurements. A number of resistivity measurements on $Fe_5GeTe_2$ reported a rapid downturn when the temperature is lowered below 100 K (*23*, *24*, *28*). While these measurements were likely conducted on samples with mixed phases, such resistivity drops are characteristic of Kondo lattices where coherent interactions between localized and itinerant bands lead to suppressed resistivity (*55*). A recent study on the flat-band phase of $Fe_5GeTe_2$ reported the emergence of a Fermi-liquid-like behavior in the low-temperature resistivity, which reinforces this Kondo-like picture (*35*). A polarized neutron scattering experiment showed that the Fe(1) atoms become magnetically ordered below 100 K in the site-disordered phase (*34*), which is inconsistent with Kondo-like behaviors. As we discussed before, the site-disordered phase should be distinguished from our Phase II which is likely associated with the "UUU" configuration of the Fe(1) atoms (*29*).

We evaluate the impact of the flat band on the tendency for Fermi surface nesting by calculating the static Lindhard response function (*36*),

$$\chi(\boldsymbol{q}) = \int \frac{d\boldsymbol{k}}{(2\pi)^d} \frac{f_{\boldsymbol{k}} - f_{\boldsymbol{k}+\boldsymbol{q}}}{\epsilon_{\boldsymbol{k}} - \epsilon_{\boldsymbol{k}+\boldsymbol{q}}} \quad (1)$$

where $\boldsymbol{k}$ and $\boldsymbol{k}$+$\boldsymbol{q}$ represent the initial and final momenta, $\epsilon_{\boldsymbol{k}}$ is the band dispersion, and $f$ is the Fermi-Dirac function. For a 2D Fermi surface associated with a dispersive band dispersion, $\chi(\boldsymbol{q})$ is maximized along the direction where the nesting vector connects parallel segments of the Fermi surface (Fig. 4A). However, from our experimentally resolved Fermi surfaces (Fig. 2A) this criterion does not lead to nesting along $\bar{\Gamma} - \bar{K}$. On the other hand, with the flat bands present at $\bar{\Gamma}$ and $\bar{K}$, $\chi(\boldsymbol{q})$ is maximally enhanced with the nesting vector connecting $\bar{\Gamma}$ and $\bar{K}$, leading to the observed $\sqrt{3} \times \sqrt{3}R30°$ charge order (Fig. 4B). Another scenario is to start with a flat band near $\bar{\Gamma}$ and a dispersive electron-like band near $\bar{K}$ (Fig. 4C), and the $\sqrt{3} \times \sqrt{3}R30°$ charge order folds the quasi-flat band from $\bar{\Gamma}$ to $\bar{K}$. Such a configuration also leads to maximally enhanced $\chi(\boldsymbol{q})$ along $\bar{\Gamma} - \bar{K}$ (Fig. 4C). Therefore, the flat band at $E_F$ plays a pivotal role in driving the electronic ordering in Phase II of $Fe_5GeTe_2$. This also explains why the band folding is restricted to 30 meV below $E_F$ (Fig. 1H and 1I), where the flat bands reside.

In summary, we reveal a distinct phase of $Fe_5GeTe_2$ in which a flat band is formed at $E_F$ and drives a $\sqrt{3} \times \sqrt{3}R30°$ charge order. Clear signatures of the band folding within 30 meV below $E_F$ are observed by high-resolution ARPES, coinciding with the energy range where the flat band resides. Taken together with our calculations of the Lindhard response function, our results suggest $Fe_5GeTe_2$ as a model system where the interaction-driven flat band directly drives the electronic order. Such a flat band may arise from Kondo-like interactions. More interestingly, the intrinsic magnetism in $Fe_5GeTe_2$ suggests that the time-reversal symmetry is spontaneously broken, leading to the theoretical possibility to realize Chern bands and topological phases (*16*). This highlights the general impact of our observations for low-energy electronic order engineering, and a pressing need to combine micro-device fabrication and low-temperature transport to probe the topological electronic properties.

## Materials and Methods

### Sample growth

Single crystals were grown via iodine-assisted chemical vapor transport following previous methods (*24*). Fe powder, Ge powder, and Te powder were weighed in the molar ratio of 5:1:2, mixed, and placed within a quartz tube along with 2.5 mg/cm$^3$ of $I_2$ pieces. The tube was then sealed under a low-pressure Ar atmosphere, and the sealed tubes were placed in a horizontal furnace using the natural temperature gradient with the source material at the center of the furnace. The furnace was ramped to 760 °C in 12 hours, dwelled for 2 weeks, and then slowly cooled to room temperature for slow-cooling crystals. Some crystals were subsequently re-sealed in a quartz tube under vacuum, placed in a furnace at 750 K for two hours, and quenched in ice water to get quenched crystals. We could also create a quenched sample by heating it for 20 minutes on a hot plate at 573 K, and quenching it by thermal contact with a metal kept at 77 K.

**ARPES measurements**

The high-resolution ARPES measurements were performed on a Multi-Resolution Photoemission Spectroscopy platform at the University of Chicago (*56*). The 6 eV laser was generated by frequency quadrupling of a Ti:Sapphire oscillator with an 80 MHz repetition rate. The 6 eV laser spot size is 10 × 15 $\mu m^2$ (FWHM) and incident on the sample at ≈45°. The analyzer slit is perpendicular to the plane defined by the laser propagation direction and the sample normal. The energy resolution of the 6 eV laser ARPES is 4 meV. The 21.2 eV ARPES measurements were carried out by using a helium discharge lamp (Scienta VUV5000) with an energy resolution of 6 meV.

**DFT + DMFT calculations**

DFT+DMFT calculations were performed using Rutgers DFT+Embedded DMFT (eDMFT) functional code (*57*). The correlation effect of Fe 3*d* orbitals is treated by a DMFT loop on top of an effective one-electron Hamiltonian obtained from the WIEN2k calculation, which employs the full-potential augmented plane wave method (*58*). Due to the computational cost, we modeled a $Fe_5GeTe_2$ monolayer for the calculations, rather than the bulk structure which contains three layers per unicell due to stacking. The five inequivalent Fe sites within the structure were treated as five different impurity problems within the DMFT framework. A local DMFT basis was constructed by diagonalizing the hybridization matrix at zero frequency in the real-harmonic representation. The hybridization energy window, spanning from -10 to 10 eV with respect to the Fermi level, was adopted for the charge self-consistent calculations. The interaction parameters were chosen as $U$ = 5 eV and $J$ = 0.8 eV, and the nominal double-counting correction was adopted. The impurity problem was solved using continuous-time quantum Monte Carlo (CTQMC) method (*59*).

## Acknowledgments

We thank Han Wu, Ming Yi, Xielin Wang, Shengxi Huang, Jianxin Zhu, and Zhi-Xun Shen for helpful discussions.

**Funding:**

The U.S. Department of Energy DE-SC0022960 (SY, QG, GB, KDN, HL)

The U.S. Department of Energy DE-SC0019443 (CH, XX, JC)

The Gordon and Betty Moore Foundation GBMF12763 (PL, SY)

**Author contributions:**

Conceptualization and supervision: SY

ARPES measurement: QG, GB, KDN, HL, SY

Data analysis: QG, GB, KDN, SY

Single crystal growth: CH, PM, XX, JC

Simulation of the static Lindhard response function: KDN, SY

First-principles calculations: BGJ

Theoretical support: PL, BGJ, BG

Writing—original draft: SY, QG, GB, KDN

Writing—review & editing: All authors

**Competing interests:** The authors declare that they have no competing interests.

**Data and materials availability:** The data that support the findings of this study are available from the corresponding author.

**Figures and Tables**

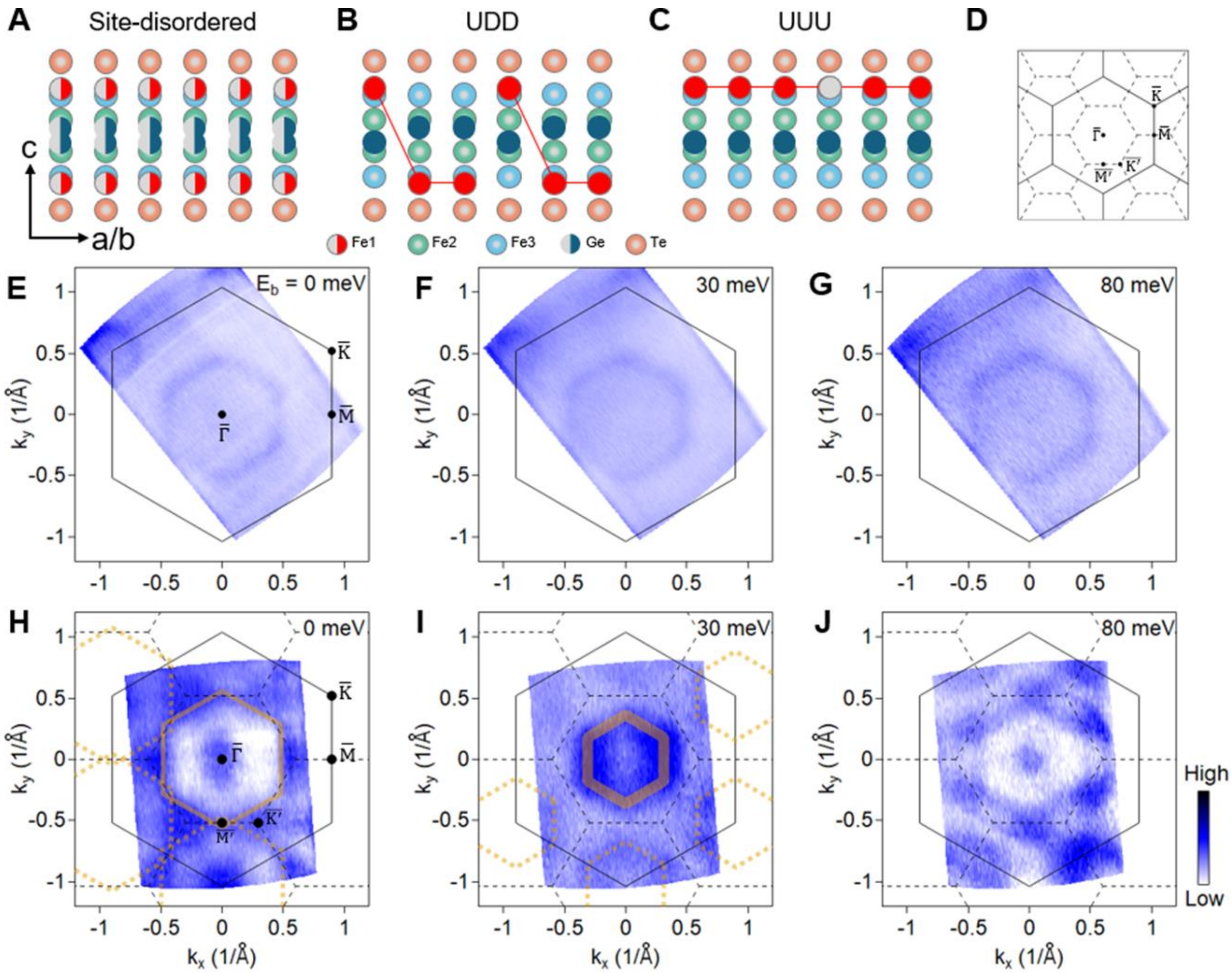

**Fig.1. Distinct phases in $Fe_5GeTe_2$.** (**A-C**) Schematic crystal structures of different $Fe_5GeTe_2$ phases, projected along the [120] direction. The Fe(1) atoms, shown as red spheres, can occupy two split-site positions-either above or below the Ge atom-referred to as the U (up) and D (down) sites, respectively. In the site-disordered phase (A), the Fe(1) atoms randomly occupy the U and D sites. In the UDD phase (B), Fe(1) atoms form an ordered superstructure with a periodic U-D-D arrangement. In the UUU phase (C), all Fe(1) atoms occupy the U sites. Half-filled circles indicate a 50% occupancy probability, and empty circles indicate vacancies. (**D**) Brillouin zone schematics: black solid lines show the original Brillouin zone, and black dashed lines show the reconstructed $\sqrt{3} \times \sqrt{3}$ Brillouin zone. (**E-G**) Fermi surface map (**E**) and constant energy contours at binding energies of 30 meV (**F**) and 80 meV (**G**) for the UDD phase. (**H-J**) Same as (**E-G**), but measured for the UUU phase. All data in this figure were measured at 8 K using unpolarized 21.2 eV photons.

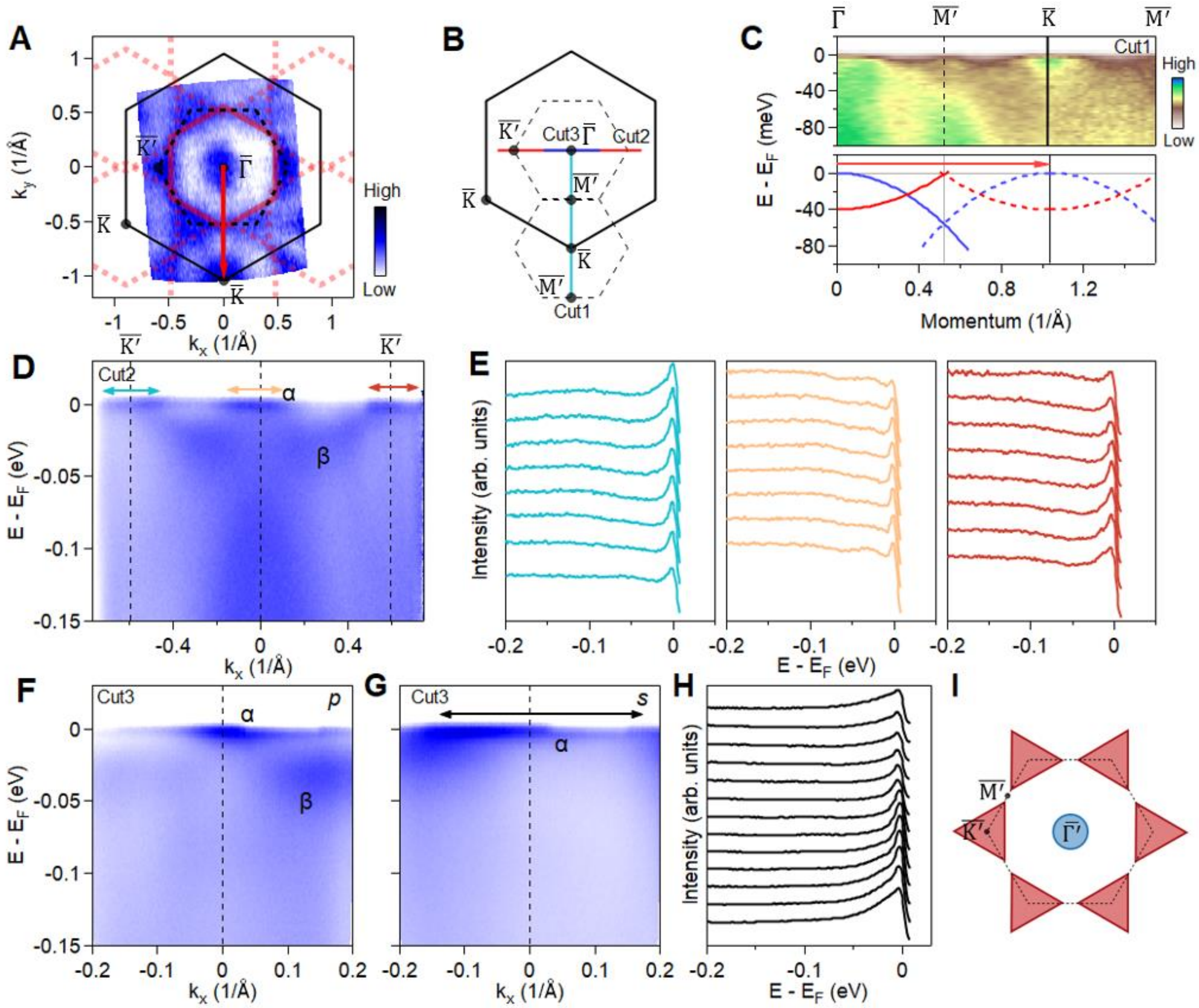

**Fig. 2. Charge order and flat band in the UUU phase of $Fe_5GeTe_2$.** (**A**) Fermi surface map. (**B**) Brillouin zone schematics: solid hexagon indicates the original Brillouin zone, and the dashed hexagons represent the reconstructed $\sqrt{3} \times \sqrt{3}$ Brillouin zone. (**C**-**D**) High-symmetry momentum cuts, with cut directions indicated in (**B**). In (C), the blue and red curves mark the electron and hole bands near $\overline{\Gamma}$, while dashed curves mark their replica bands folded to $\overline{K}$ due to the $\sqrt{3} \times \sqrt{3}$ reconstruction. (**E**) Momentum-dependent energy-distribution curves (EDCs) taken at equally spaced momentum points in D. (**F**-**G**) Dispersions along Cut3 measured with 6-eV laser in *p* (F) and *s* (G) polarizations. (**H**) EDCs taken from (**G**), emphasizing the flat band at the Fermi level. (**I**) Schematic of the reconstructed Fermi surface in the reconstructed Brillouin zone. The blue circle near $\overline{\Gamma'}$ corresponds to the flat band connected to a hole-like band, while red triangles near $\overline{K'}$ indicate flat bands connected to an electron-like band (fig. S6). Data in (**D**-**H**) are divided by the Fermi-Dirac function convolved with the energy resolution.

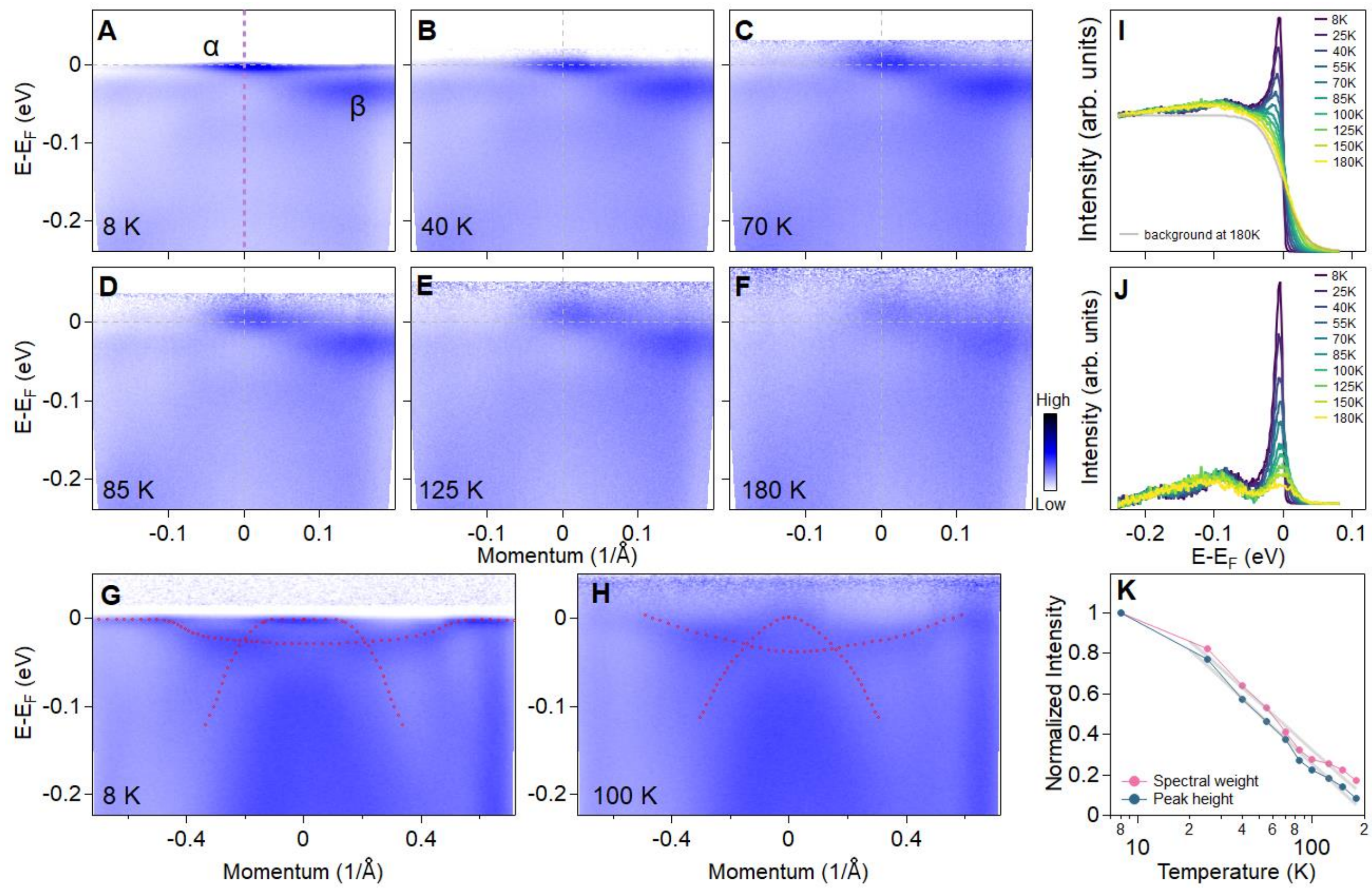

**Fig. 3. Temperature-dependent flat band in the UUU phase of $Fe_5GeTe_2$.** (**A**-**F**) ARPES spectra along $\bar{\Gamma} - \bar{M}$ at various temperatures from 8 K to 180 K, taken with a 6 eV *p*-polarized laser. (**G, H**) ARPES spectra along $\bar{\Gamma} - \bar{M}$ at 8 K (**G**) and 180 K (**H**), acquired with 21.2 eV photons. All spectra in (**A**-**H**) are divided by the Fermi-Dirac function convolved with the energy resolution. (**I**) Temperature-dependent energy-distribution curves (EDCs) at $\bar{\Gamma}$, showing the formation of the coherent peak. No Fermi-Dirac division is applied here. (**J**) EDCs from (**I**) after subtraction of a Fermi-Dirac distribution background. The 180 K background used for the subtraction is indicated by the grey line in (**I**). (**K**) Temperature dependence of the spectral weight (pink dotted line) and coherent-peak height (blue dotted line) extracted from (**J**) and normalized to their values at 8 K. The grey lines in (K) are linear fits of the data as a function of log(T). The spectral weight is obtained by integrating the EDCs from -0.05 to 0 eV.

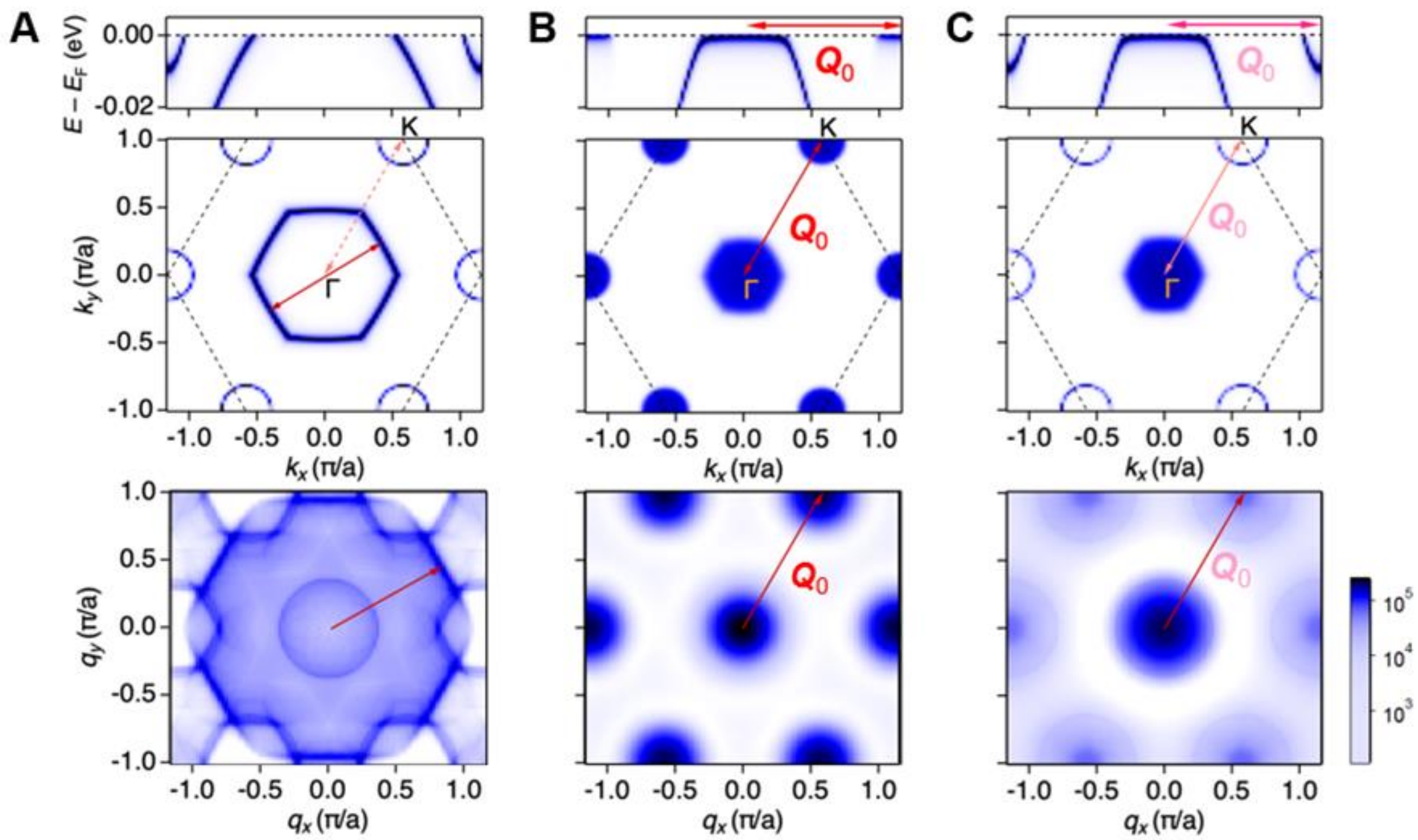

**Fig. 4. Theoretical picture of flat-band-induced nesting in $Fe_5GeTe_2$.** (**A**) Simulated static Lindhard response function (bottom) for a simplified dispersive band structure (top), with the corresponding Fermi surface map (middle) showing a hole pocket near $\bar{\Gamma}$ and an electron pocket near $\bar{K}$. This simulation shows preferred nesting vectors that connect the 2 opposite sides of the hole pocket at $\bar{\Gamma}$. (**B**) A similar simulation with flat bands near $E_F$ at $\bar{\Gamma}$ and $\bar{K}$. (**C**) A similar simulation with a flat band near $E_F$ at $\bar{\Gamma}$ and a dispersive electron pocket near $\bar{K}$.

# Supplementary Materials for

**Interaction-driven flat bands and charge order in $Fe_5GeTe_2$**

Qiang Gao et al.

*Corresponding author: Shuolong Yang. Email: yangsl@uchicago.edu

This PDF file includes:

Figs. S1 to S10
Supplementary text

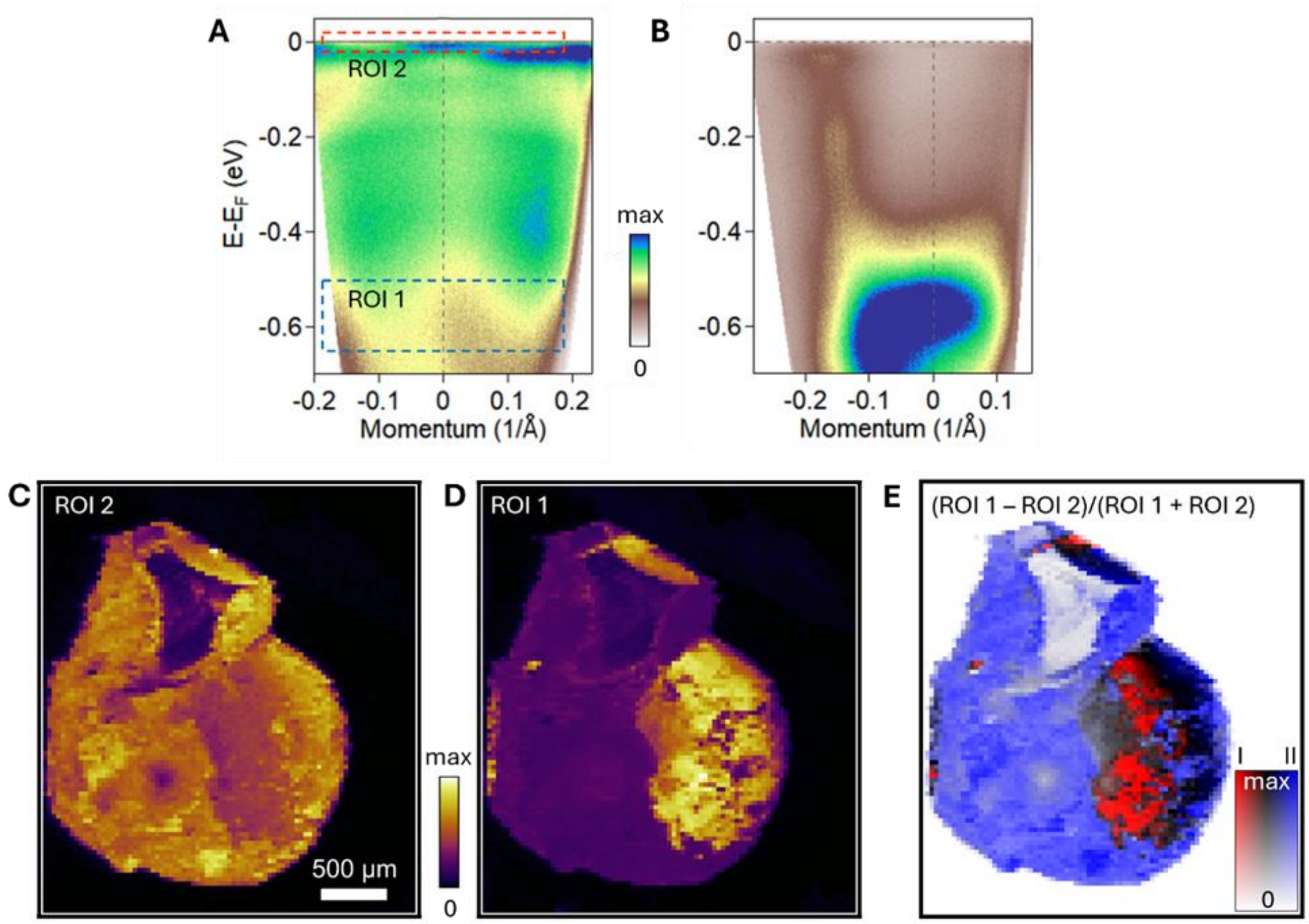

**Fig. S1**: **Micro-ARPES maps of flat-band and site-ordered domains.** (**A**) Typical 6 eV ARPES spectrum on a flat-band area (Phase II in the main text). (**B**) Same, for a site-ordered area (Phase I in the main text). (**C**, **D**) ARPES spatial maps of a $Fe_5GeTe_2$ sample, where the intensity corresponds to the spectral weight integrated in the Region of Interest (ROI) 2 (**C**) and ROI 1 (**D**) as indicated in (**A**), highlighting the flat bands near $E_{\mathrm{F}}$ and the -0.6 eV peak, respectively. It is easy to distinguish the different domains. Here $\mathrm{ROI}\ 1 = ([-20{:}\ 20]\ \ \mathrm{meV}) \times ([-0.19{:}\ 0.19]\ 1/\text{Å})$ and $\mathrm{ROI}\ 2 = ([-500{:} -650]\ \ \mathrm{meV}) \times ([-0.19{:}\ 0.19]\ 1/\text{Å})$ . (E) Differential image (dichroism) between (D) and (C), highlighting the two domains: red for Phase I and blue for Phase II. All data were acquired at 8 K.

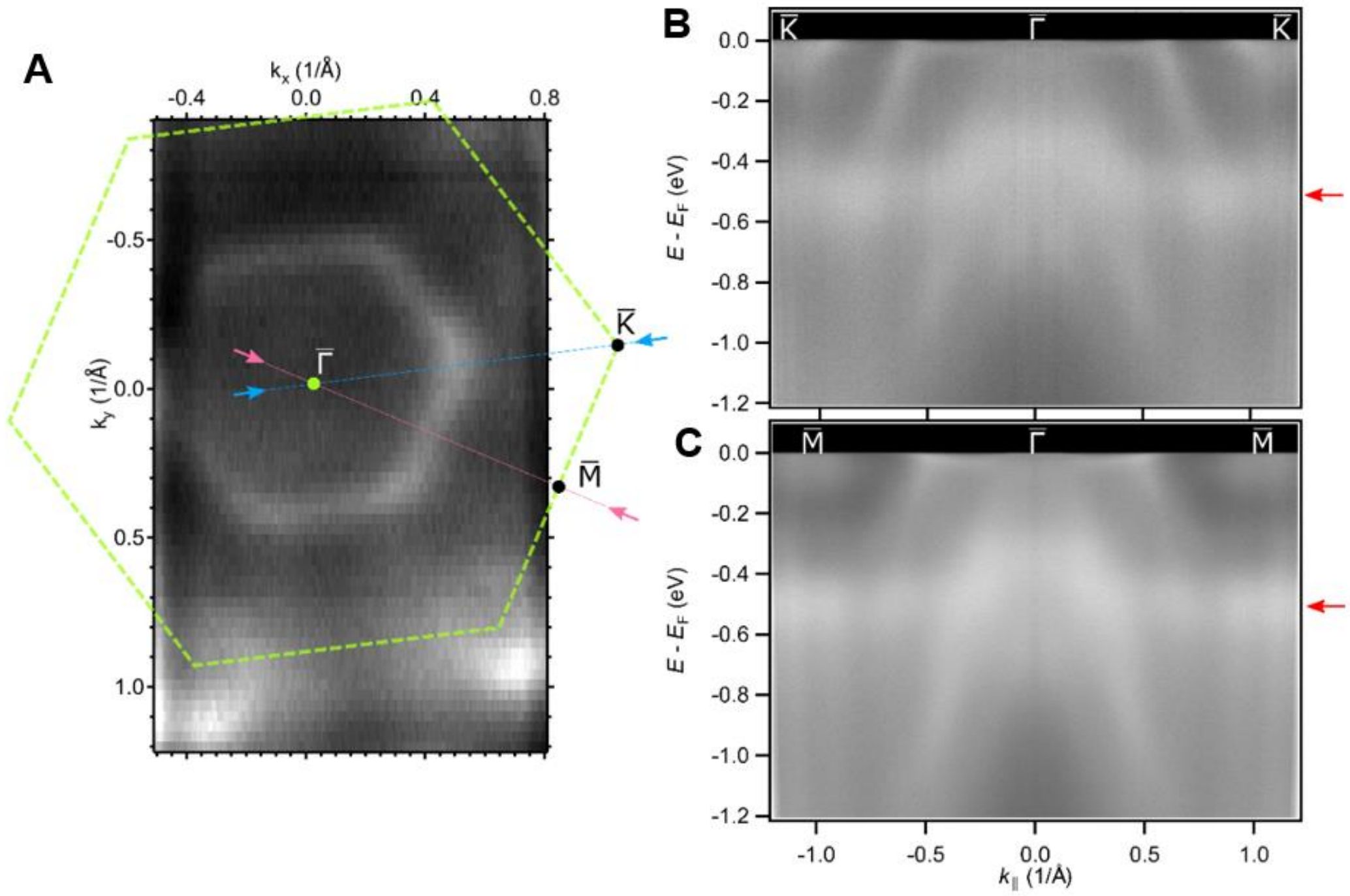

**Fig. S2**: **21.2 eV ARPES measurements of the site-ordered phase (Phase I in the main text).** (**A**) Fermi-surface map and sketch of the first Brillouin zone boundaries (dashed green line). (**B**) Single cut along the $\bar{\Gamma} - \bar{K}$ direction. (**C**) Same, for the $\bar{\Gamma} - \bar{M}$ direction. The flat band at ≈ −0.5 eV (red arrows in (**B**) and (**C**)) extends over a large portion of the Brillouin zone away from $\bar{\Gamma}$. These measurements were performed at T = 8 K, with 21.2 eV unpolarized photons on a ≈1.5×1.5 mm² site-ordered cleave which contained <5−10% of other phases. The measured spectra include a faint signal from the flat-band phase, mostly around $\bar{\Gamma}$ and $E_F$. The spectra (**B**, **C**) have been measured along the blue and pink solid lines in (**A**), respectively, and symmetrized around $\bar{\Gamma}$ for better data visualization.

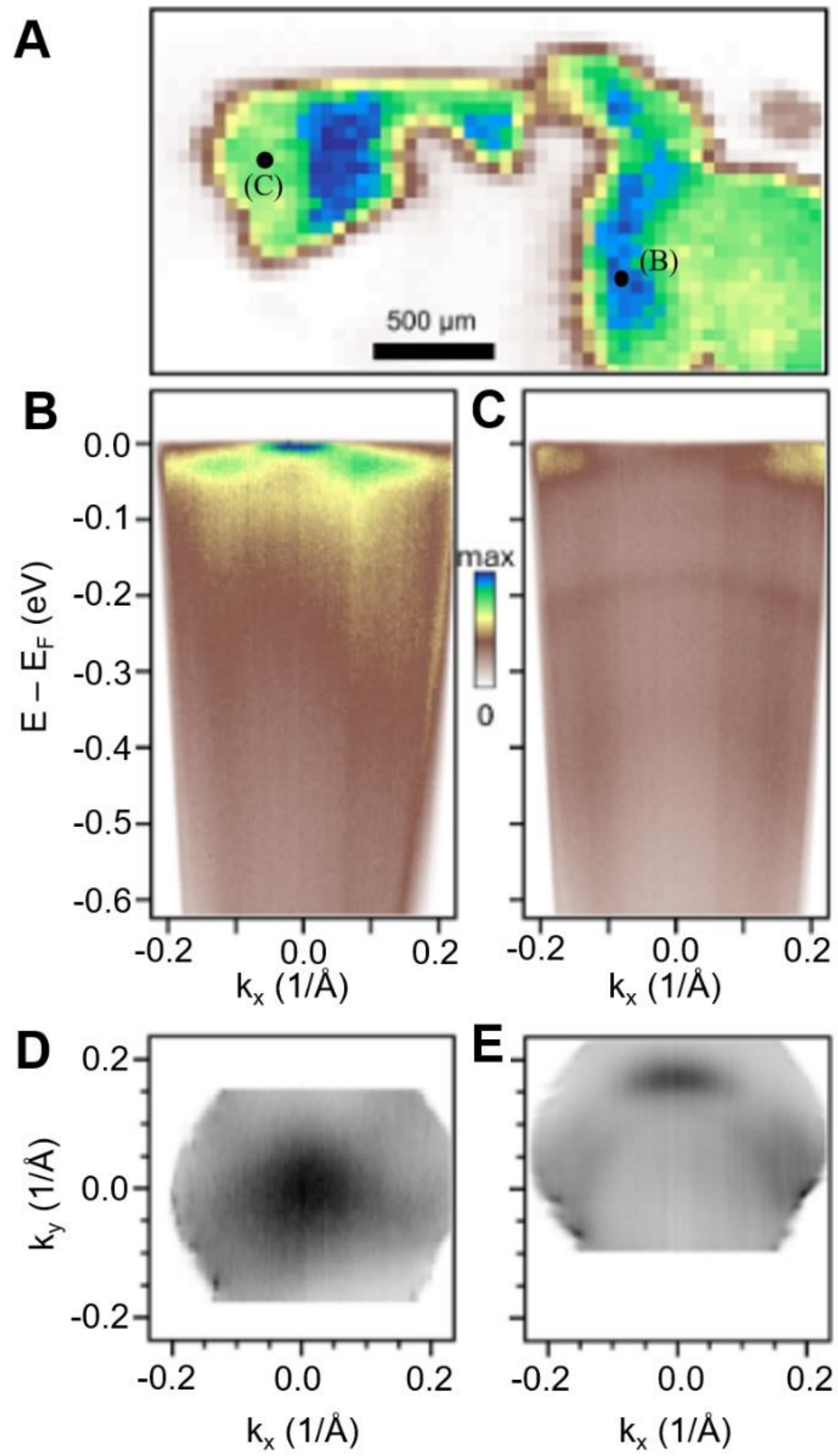

**Fig. S3**: **Different phases seen by 6 eV ARPES besides the site-ordered phase.** (**A**) ARPES micrograph of a sample cleave that does not contain site-ordered regions (Phase I in the main text). The counts are integrated over the ROI $= ([-50:20]\ \mathrm{meV}) \times ([-0.22:0.22]\ 1/\text{Å})$ (**B**, **C**) ARPES spectra for positions (B) and (C) marked in (**A**), taken along the $\bar{\Gamma} - \bar{\mathrm{K}}$ direction. (**D**, **E**) Corresponding Fermi surface maps of the two phases. All the blue-colored regions in (**A**) share the same flat-band phase of panel (**B**), referred to as Phase II in the main text. The green/yellow regions present a different phase, with distinct band structure and Fermi surface, as shown in (**C**) and (**E**). All spectra were acquired at 8 K.

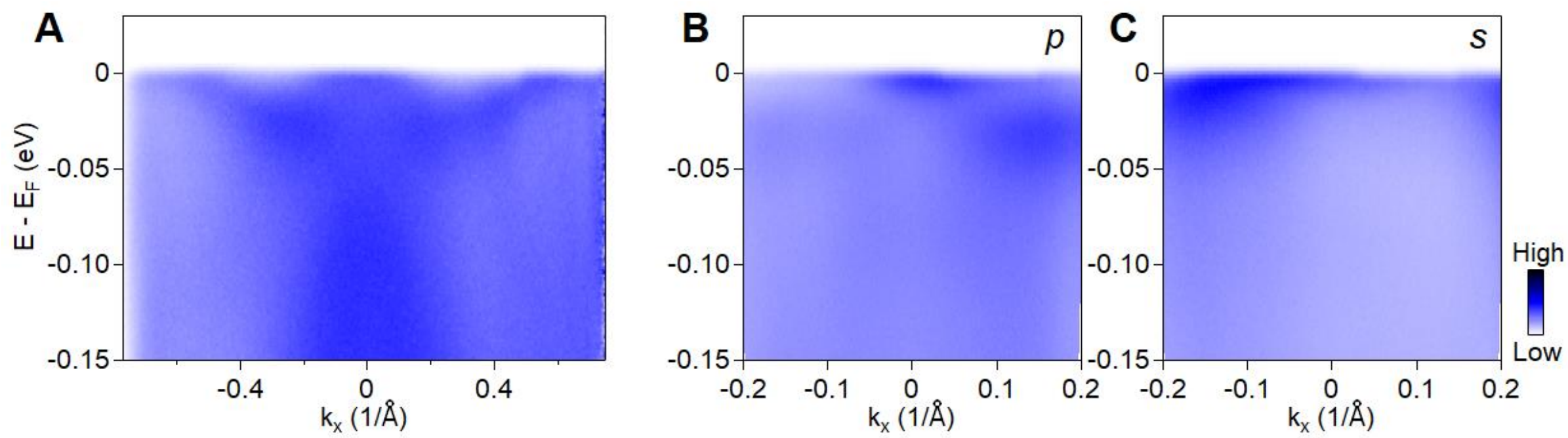

**Fig. S4: The band structure of the UUU phase.** (**A**) The raw spectra corresponding to Fig. 2**D** in the main text. (**B**) corresponds to Fig. 2**F**. (**C**) corresponds to Fig. 2**G**.

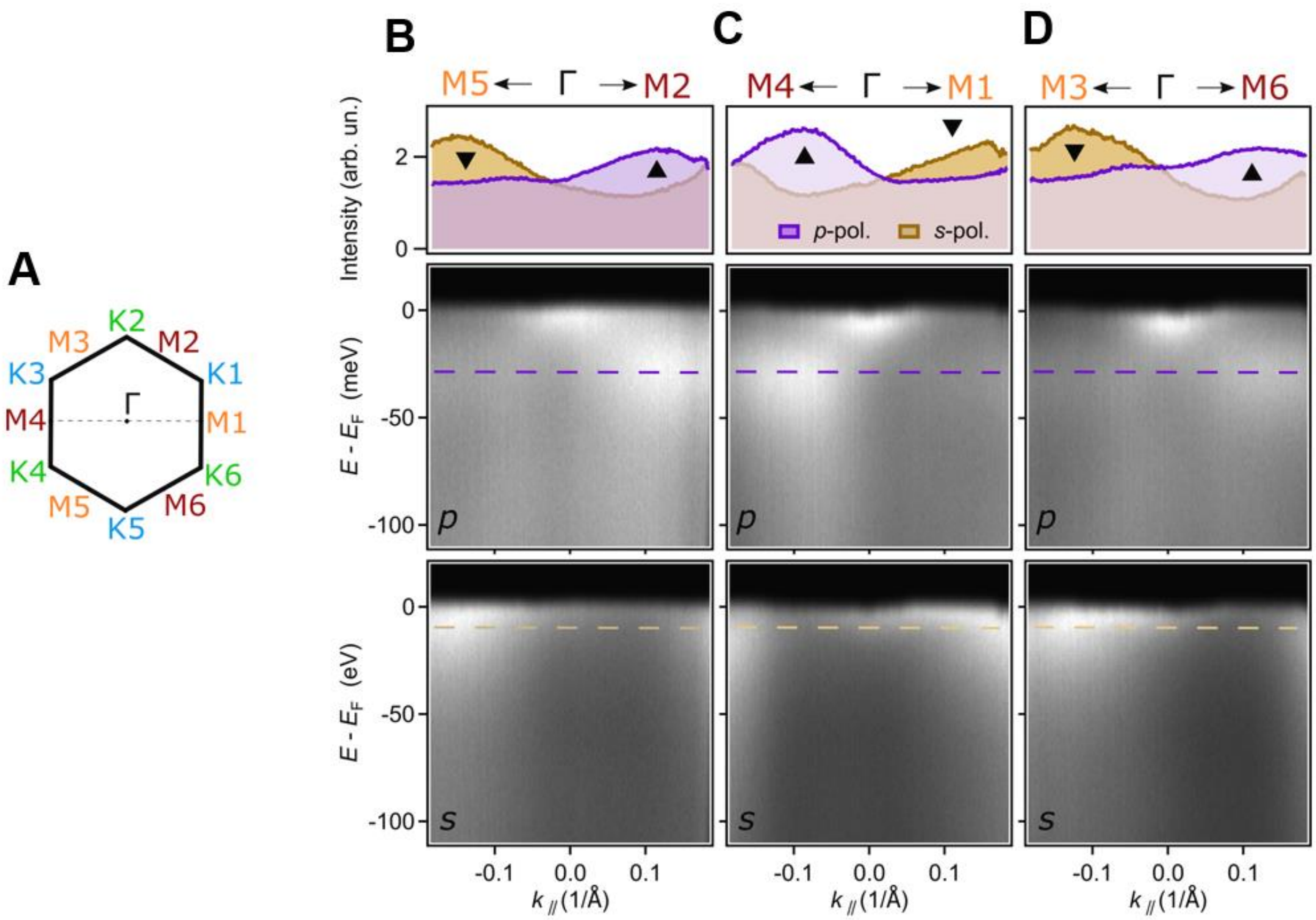

**FIG. S5: $C_3$ symmetry of the electronic structure.** (**A**) Sketch of the Brillouin zone and its high-symmetry points. Under the $C_3$ symmetry, M1, M3, and M5 are equivalent; M2, M4, and M6 are equivalent. The M points with odd and even indices are not equivalent. The same holds for the K points. (**B**-**D**) ARPES spectra measured along the three M-Γ-M cuts for *p* and *s* polarizations. M5-Γ-M2 is equivalent to M3-Γ-M6 but not equivalent to M4-Γ-M1. The curves on top are momentum distribution curves (MDCs) at $E = -35$ meV and $E = -10$ meV for *p* and *s* polarizations, respectively. These measurements were performed at T = 8 K with 6 eV linearly polarized photons on a Phase II domain of $Fe_5GeTe_2$.

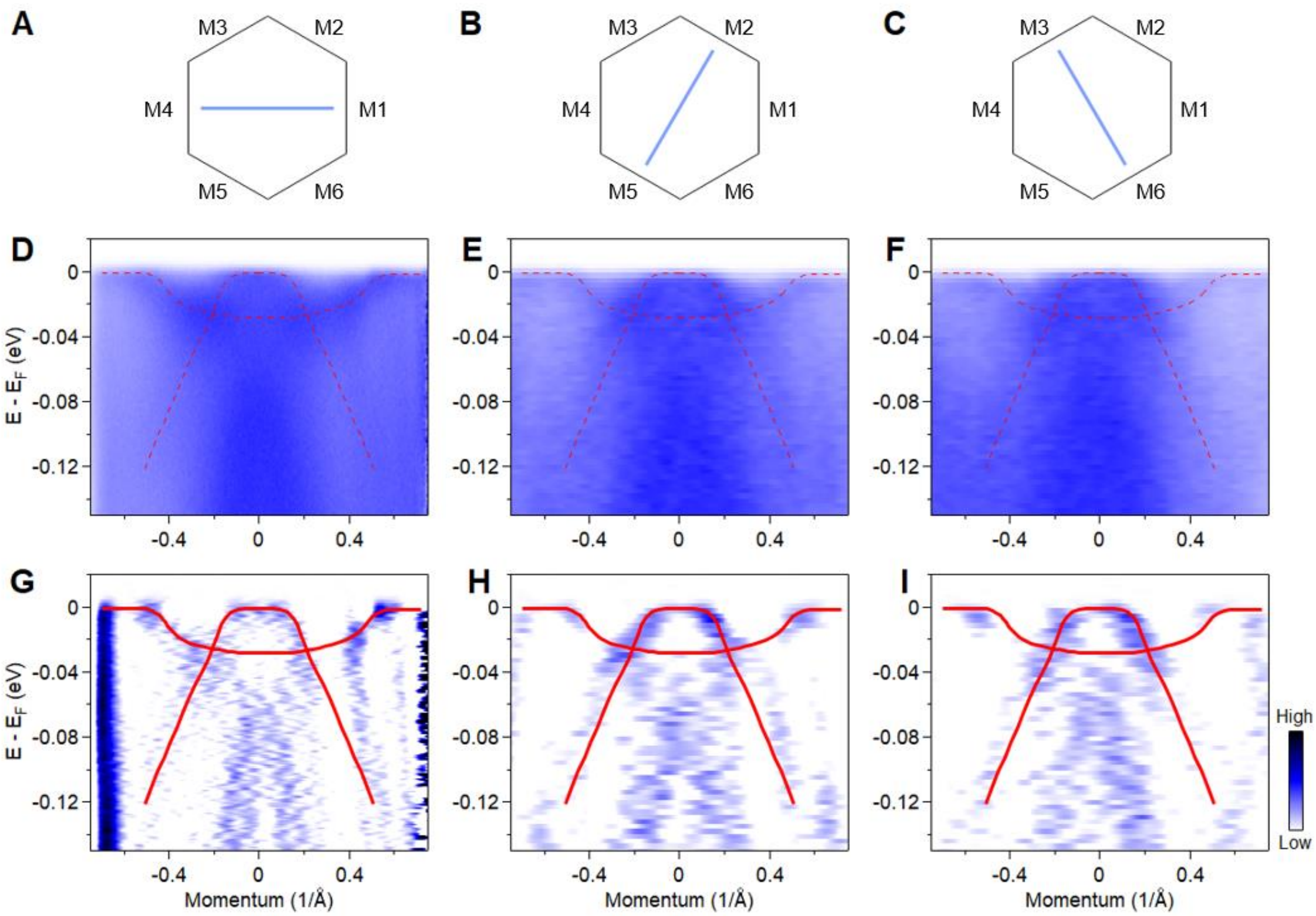

**FIG. S6: The electronic structure of the UUU phase.** (**A-C**) Sketch of the Brillouin zone and three high-symmetry cuts. (**D-F**) ARPES spectra measured along the three M-$\overline{\Gamma}$ -M cuts marked in (**A-C**), taken with 21.2 eV photons at 8 K. (**G-I**) The second derivative image of (**D-F**) along the momentum axis. The red curves in (**D-G**) mark the electron and hole bands.

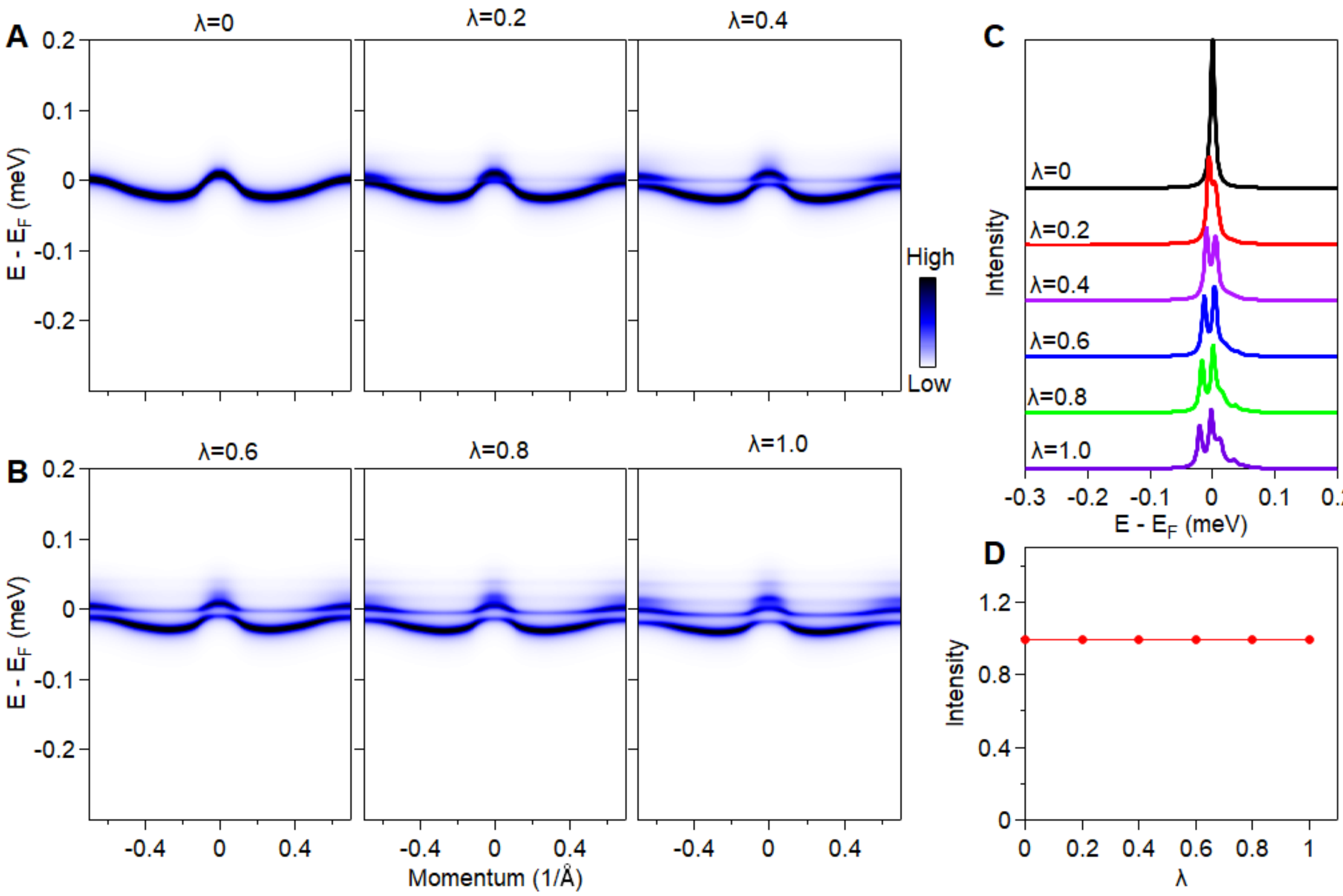

**FIG. S7: Spectral function simulation of Holstein polarons.** (**A**, **B**) Calculated spectral functions with different electron-phonon coupling constants λ. (**C**) Energy distribution curves at $\bar{\Gamma}$ for the different λ's. (**D**) The integrated spectral weight for the different λ's taken from **C**.

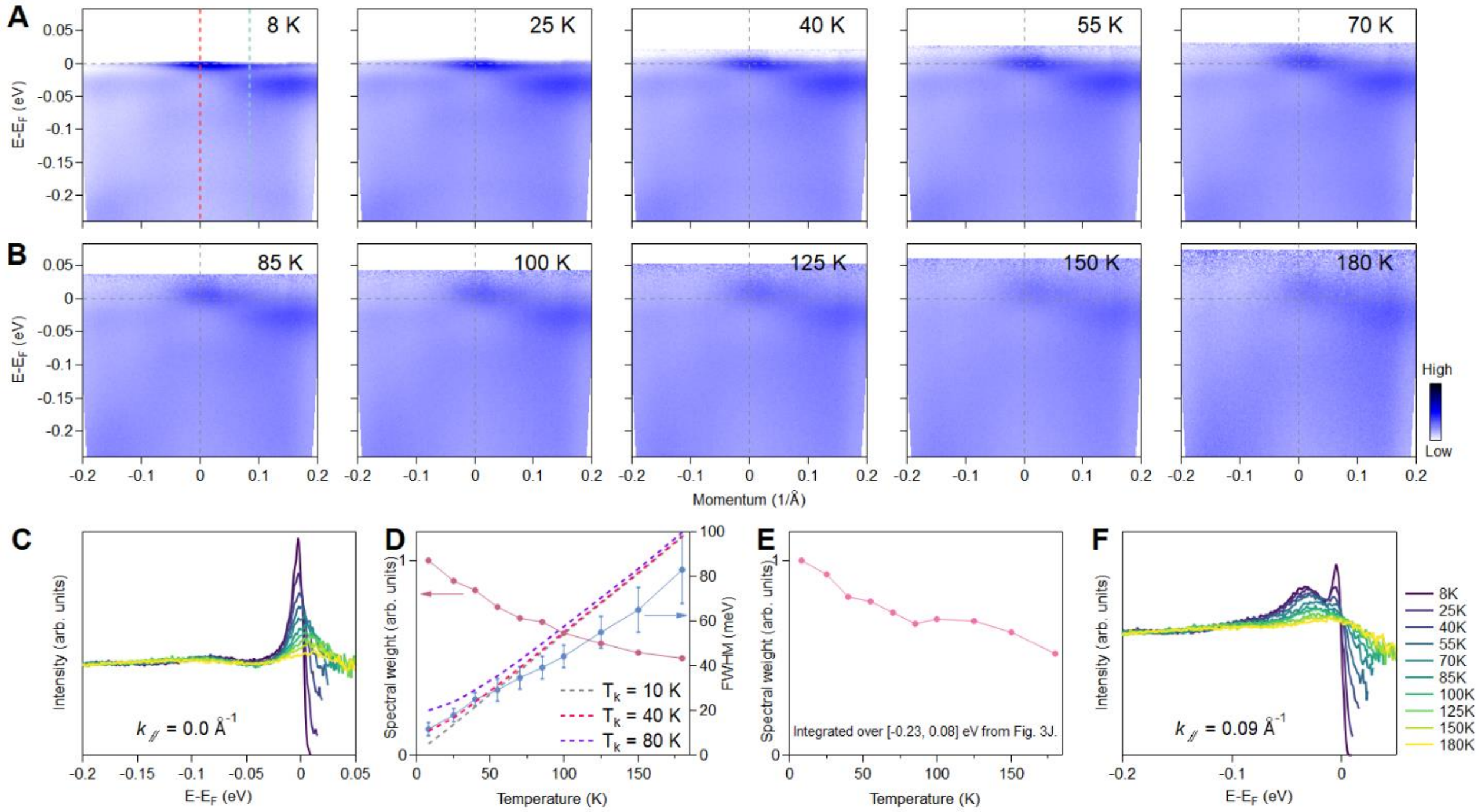

**FIG. S8: Additional data and analysis for the temperature dependence of the flat band.** (**A-B**) ARPES spectra along $\bar{\Gamma} - \bar{M}$ at various temperatures from 8 K to 180 K, taken with a 6 eV *p*-polarized laser. (**C**) Temperature-dependent energy-distribution curves (EDCs) taken at $\bar{\Gamma}$, showing the formation of the coherent peaks. The EDCs are divided by the Fermi-Dirac function convolved with the energy resolution. (**D**) Temperature dependence of the spectral weight (purple dotted line) and full width at half maximum (FWHM, blue dotted line) obtained by fitting the coherent peaks of the EDCs in (**C**) to Lorentzian functions. Dashed curves are the calculated temperature dependence of an effective Kondo resonance peak width with assumed single-ion Kondo temperatures $T_K$ = 10, 40, and 80 K. (**E**) Temperature dependence of the spectral weight, obtained by integrating the EDCs in Fig. 3J in the main text from -0.23 to 0.08 eV. (**F**) Temperature-dependent EDCs taken at a momentum of $0.09\ \text{Å}^{-1}$. The momentum position is indicated by the blue dashed line in (**A**). The EDCs are divided by the Fermi-Dirac function convolved with the energy resolution.

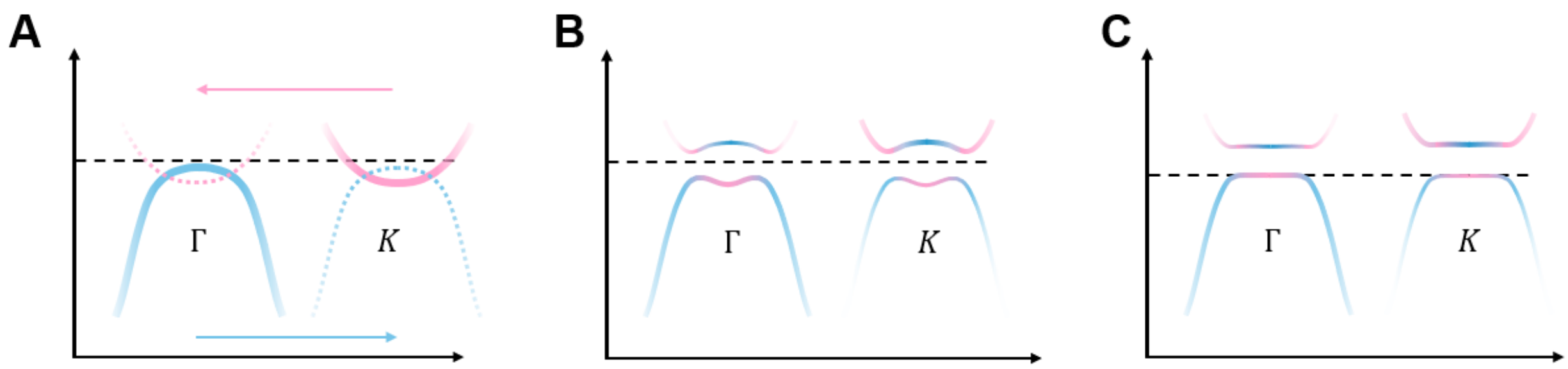

**FIG. S9: Schematic illustration of excitonic flat bands.** (**A**) The band structure of a system with a hole band at $\bar{\Gamma}$ and an electron band at $\bar{K}$. The charge order connecting $\bar{\Gamma}$ and $\bar{K}$ generates backfolded replicas of the electron band at $\bar{\Gamma}$ and of the hole band at $\bar{K}$. (**B**) The hybridization between the original and backfolded bands yields "M" shaped dispersions at both $\bar{\Gamma}$ and $\bar{K}$. (**C**) Due to the strong correlation effect, the band top can become flat and moves to the Fermi level.

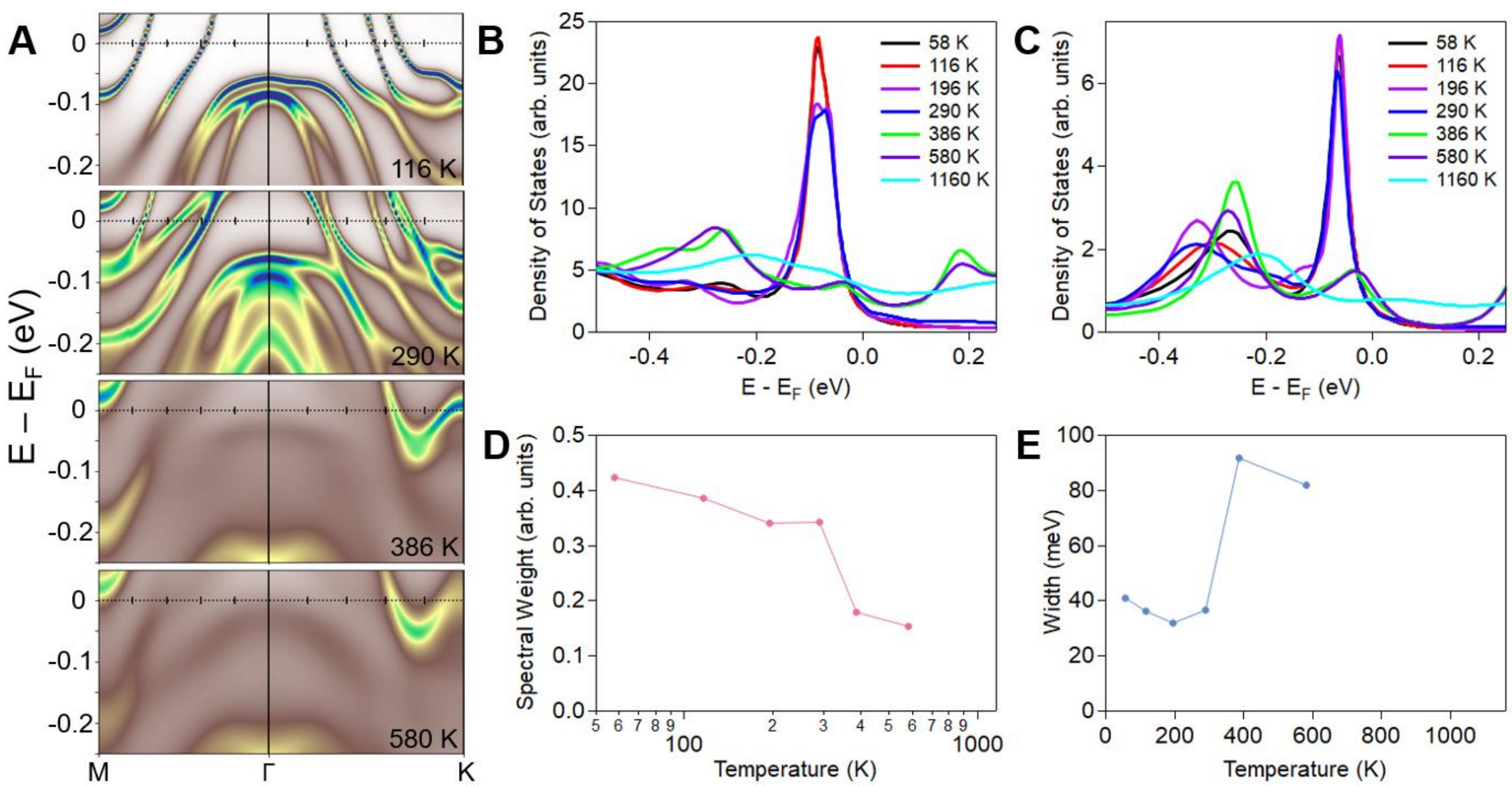

**Fig. S10: Temperature-dependent electronic structure of $Fe_5GeTe_2$ in the UUU phase from DFT + DMFT.** (**A**) Calculated spectral function at selected temperatures along M-Γ-K. Multiple hole-like bands with relatively flat tops are observed near Γ at energies of approximately -0.06 eV and -0.1 eV. (**B**) Temperature evolution of the momentum-integrated density of states near Γ extracted from the calculated band structure. (**C**) Temperature dependence of the density of states projected onto selected orbitals that contribute to the flat band around -0.06 eV. (**D**, **E**) Temperature dependence of the spectral weight and linewidth of the -0.06 eV feature extracted from (**C**).

**1, Discussion on the phase assignment.**

We address the significant controversies in the ARPES literature of $Fe_5GeTe_2$ due to the multiple structural configurations of $Fe_5GeTe_2$ and the complexity of its band structure (*28*, *30*-*35*). Earlier ARPES studies on $Fe_5GeTe_2$ did not explicitly distinguish different structural configurations (*30*, 31, *33*). Even though some studies (*31*, *33*) claimed to have observed the band folding between $\overline{\Gamma}$ and $\overline{\mathrm{K}}$, the replication of the Fermi surfaces was not clearly observed. Later ARPES studies (*28*, *32, 34, 35*) discovered that there are multiple structural configurations leading to distinct electronic structures. Wu *et al.* (*28*) identified the electronic structures for the site-ordered ("UDD" or "DUU") and site-disordered phases. Huang *et al.* (*32*) assigned two types of electronic structures to the Fe(1) and Fe(5) terminations. Our Phase I exhibits an electronic structure that mimics both the site-ordered phase in Wu *et al.* and the Fe(5) termination in Huang *et al.*: dispersive hole pockets near $\overline{\Gamma}$ and a small electron pocket near $\overline{\mathrm{K}}$. Considering that Wu *et al.* also conducted optical second harmonic generation measurements to confirm the symmetry, we followed this study to assign our Phase I to the site-ordered phase. Importantly, our Phase II featuring diffusive features in the Fermi surface map appears different from either the site-disordered phase in Wu *et al.* (*28*) or the Fe(1) termination in Huang *et al.* (*32*). Instead, our observations of the clear $\sqrt{3} \times \sqrt{3}$ reconstruction and of the flat band at $E_F$ are more consistent with the STM study on the "UUU" configuration with a charge order (29). This STM study also showed that the site-ordered phase possesses nanoscale mixtures of "UDD" and "DUU" domains, which explains why no long-range coherent $\sqrt{3} \times \sqrt{3}$ reconstruction due to the lattice order can be observed in ARPES (*28*). In comparison with another recent ARPES study (*35*), our work agrees on the attribution of the quenched phases to the UDD or DUU configurations, although we cannot separately resolve electronic structures corresponding to these two phases. We point out that there is a strong similarity between our Phase II and the claimed Kondo lattice phase in Huh *et al*., but the $\sqrt{3} \times \sqrt{3}$ Fermi surface reconstruction was not reported in Huh *et al.* This could be due to improved sample quality or different photoemission matrix elements associated with our 6 eV photons as compared to those with the 7 eV photons used in Huh *et al*.

**2, Simulation of the spectral function of Holstein polarons.**

The spectral function of Holstein polarons is calculated by taking the analytical approximation of the Green's function

$$G(k,\omega) = \frac{1}{\omega - \varepsilon_k - \Sigma_{MA}(\omega) + i\eta} \quad (1)$$

where $\omega$ is the energy, $\epsilon_k$ is the bare band dispersion extracted from the experimental data at high temperature, $\eta$ is the spectral broadening, and $\Sigma_{MA}$ is a momentum-averaged self-energy with form (*60*):

$$\Sigma_{MA}(\omega) = \cfrac{g^2\,\overline{g}_0(\omega - \Omega_0)}{1 - \cfrac{2\,g^2\overline{g}_0(\omega - \Omega_0)\,\overline{g}_0(\omega - 2\Omega_0)}{1 - \cfrac{3g^2\,\overline{g}_0(\omega - 2\,\Omega_0)\,\overline{g}_0(\omega - 3\Omega_0)}{1 - \cdots}}} \quad (2)$$

$$\overline{g}_0(\omega) = \frac{sgn(\omega)}{\sqrt{(\omega + i\eta)^2 - 4t^2}} \quad (3)$$

where $\Omega_0$ is the phonon energy taken from the experimental data as 15 meV, $g^2 = 2t\Omega_0\lambda$ for 1D model, $t$ is the hopping amplitude and $\lambda$ is the electron-phonon coupling constant. Figure S7 shows the calculated single-particle spectral function with different $\lambda$'s.

3, **The electronic structure of $Fe_5GeTe_2$ in the UUU phase obtained from DFT + DMFT calculations.**

We performed DFT + DMFT calculations of the UUU phase using interaction parameters $U = 5$ eV and $J = 0.8$ eV. The calculated spectral function reveals multiple hole-like bands near Γ, including relatively flat band tops close to the Fermi level (fig. S10A). With increasing temperature, the flat bands become incoherent. Notably, the bands near -0.1 eV smear out faster than the flat band at -0.06 eV. To compare with experimental results, we extract the temperature-dependent density of states near Γ from the calculated spectra (fig. S10B, C). The fitted spectral weight (fig. S10D) does not follow a log($T$) temperature scaling, and the linewidth (fig. S10E) exhibits a non-monotonic temperature dependence. Therefore, although the present calculations capture flat-like bands near the Fermi level, we do not find clear evidence that the flat feature is directly connected to the Kondo-like phenomenon observed in the experiment. This suggests that the experimentally observed behavior may involve additional ingredients beyond the local correlations captured in our bulk single site DMFT calculation, such as Fe(1) defect or disorder. In particular, the observed charge-ordering phenomena imply that nonlocal electronic interactions may play an important role in shaping the low-energy spectral features. Clarifying the relative importance of these effects would require theoretical approaches beyond the current bulk DFT + DMFT treatment.